\documentclass[showpacs,aps]{revtex4}
\usepackage{graphicx}
\begin{document}

\title{On the center-vortex baryonic area law}
\author{John M. Cornwall\footnote{Email:  cornwall@physics.ucla.edu}}
\affiliation{Department of Physics and Astronomy, University of California, Los Angeles, Los Angeles Ca 90095}

\begin{abstract}
\pacs{11.15-q,12.38Aw,12.38Lg \hfill UCLA/03/TEP/16}
We correct an unfortunate error in an earlier work of the author, and show that 
in center-vortex QCD (gauge group $SU(3)$) the baryonic area law is the so-called $Y$ law, described by a minimal area with three surfaces spanning the three quark world lines and meeting at a central Steiner line joining the two common meeting points of the world lines.  (The earlier claim was that this area law was a so-called $\Delta$ law, involving three extremal areas spanning the three pairs of quark world lines.)   We give a preliminary discussion of the extension of these results to $SU(N),\;N>3$. These results are based on the (correct) baryonic Stokes' theorem given in the earlier work claiming a $\Delta$ law. The $Y$-form area law for $SU(3)$ is in agreement with the most recent lattice calculations.
\end{abstract}

\maketitle

\section{Introduction}

\begin{figure}
\includegraphics[height=1.8in]{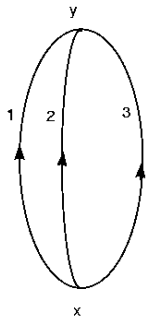}
\caption{\label{bwlpic} The $SU(3)$ baryonic Wilson loop.  [From \cite{co96}.]}
\end{figure}

In this brief report we correct an error in an earlier work of the author \cite{co96}, claiming that in center-vortex QCD (that is, for gauge group $SU(3)$) the area law for the $SU(3)$ baryonic Wilson loop (Fig. \ref{bwlpic}) involved the sum of three extremal areas, each spanning a pair of quark world lines.  We call this sort of area law a $\Delta$ law, and the corresponding area $A(\Delta )$.  We now claim that the correct law is a $Y$ law, with an extremal area $A(Y)$ composed of three areas spanning the quark world lines and a central Steiner line.  (A Steiner line is a line where three surfaces meet at angles, generically $2\pi /3$, chosen to minimize the total area of the three surfaces.)  Either the $\Delta$ law  or the $Y$ law accords with the principle, derived \cite{co03} from center-vortex theory, that area laws in general involve extremal areas.  The $Y$ law leads to an absolute minimum of the area spanning three fixed quark world lines, but the $Y$ law always has \cite{co96} higher action than that of the $\Delta$ law, because in $SU(3)$ the areas have different coefficients in the VEV $\langle W \rangle$ of the Wilson loop :
\begin{equation}
\label{ydelta}
\langle W \rangle \sim \exp [\frac{-K_F A(\Delta )}{2}]\;{\rm or}\;\exp [-K_FA(Y)].
\end{equation}
It is also worth noting that, for group-theoretical reasons,   lowest-order perturbative gluon exchange leads to (short-range) forces solely of the $\Delta$ pattern \cite{co77}. 

Recent lattice simulations of the heavy-quark baryonic area law, as far as the author knows, are in accord with the $Y$ law \cite{suga,afj,ibss}.  (Some of these workers had also found \cite{aft} evidence for a $\Delta$ law, but this work is apparently superseded by that in \cite{afj}.)  The numerical distinction between the $Y$ law and the $\Delta$ law is small, and so high accuracy is needed; it could be that the last word in lattice simulations has not yet been said.  Still, the present lattice evidence seems to favor the $Y$ law.

Section \ref{su3} points out the error in \cite{co96} and corrects it for $SU(3)$, using the baryonic Stokes' theorem of \cite{co96}.  Section \ref{sun} extends this work to $SU(N)$.  Section \ref{finis} contains conclusions. 

\section{The baryonic area law for  SU(3)}
\label{su3}

The correct parts of the arguments of \cite{co96} will be briefly summarized, and the error of this reference corrected.
The baryonic Wilson loop $W$ is:
\begin{equation}
\label{loop}
W=\frac{1}{6}\epsilon_{abc}\epsilon_{a'b'c'}U(x,y;1)_{aa'}U(x,y;2)_{bb'}U(x,y;3)_{cc'}
\end{equation}
where each $U$ is an ordered path integral from the point $x$ where the three quark world lines originate to the point $y$ where they rejoin:
\begin{equation}
\label{udef}
U(x,y;j)_{kk'}=[P\exp \int_{\Gamma(j)}dz^{\alpha}A_{\alpha}(z)]_{kk'}.
\end{equation}
Here $\Gamma(j)$ is the path from $x$ to $x$ along quark world line $j$, and we use the standard anti-Hermitean matrix representation of the gauge potential, which has the gauge coupling incorporated:
\begin{equation}
\label{adef}
A_{\alpha}(z)=\frac{g\lambda_j}{2i}A_{\alpha}^j(z).
\end{equation}
This Wilson  loop is to be averaged over all center-vortex gauge-potential configurations.  We idealize the center vortices to random closed two-surfaces ($d=4$) or closed lines ($d=3$) of infinitesimal thickness, finite persistence length, and finite density per unit area.  (In reality the vortices have finite thickness; this gives rise only to perimeter-law corrections.)  For $SU(3)$ there is only one kind of vortex, plus its anti-vortex; these can be distinguished by assigning an orientation to the closed surface (line) representing the vortex in $d=4(3)$.  This orientation can be considered as equivalent to the direction of Abelian magnetic field lines lying in the vortex surface (line), as indicated in the figures.  Just as in \cite{co03} this leads to the argument that the VEV $\langle W \rangle$ has a form as in Eq. (\ref{ydelta}) which is exponential in an extremal area.  The extremal area for the $\Delta$ law is the sum of three areas, each extremal for the three pairs of quark lines that it spans, and for the $Y$ law it is the absolute minimum area spanning the three lines jointly.

For an infinitesimally-thick center vortex, we can choose the corresponding gauge potential as a pure singular gauge:
\begin{equation}
\label{vortex}
A_{\alpha}(z)=V(z)\partial_{\alpha}V^{-1}(z).
\end{equation}
In particular, up to a regular gauge transformation $V$ can be chosen to be of the form 
\begin{equation}
\label{vform}
V(z)=\exp [2\pi iQ\beta (z)]
\end{equation}
where $Q$ is the diagonal matrix (1/3,1/3,-2/3) and $\beta (z)$ gives rise to an Abelian gauge potential:
\begin{equation}
\label{abelpot}
\tilde{A}_{\alpha}(z)=\partial_{\alpha}\beta (z)
\end{equation} 
that is singular on a closed surface and has unit magnetic flux.  For the anti-vortex, $Q\rightarrow -Q$.

For this pure gauge potential the path integrals of Eq. (\ref{udef}) are:
\begin{equation}
\label{gaugeu}
U(x,y;j)=V(x)V^{-1}(y)\equiv \exp [2\pi iQ\Lambda (j)];\;\Lambda (j)=\int_{\Gamma(j)}dz^{\alpha}\tilde{A}_{\alpha}(z).
\end{equation} 
This path integral may or may not depend on the path $j$; if it does not, then the Wilson loop of Eq. (\ref{loop}) is just $\det U=1$. If it does, then the vortex is linked to the Wilson loop and further analysis is needed.  

\begin{figure}
\includegraphics[width=3in]{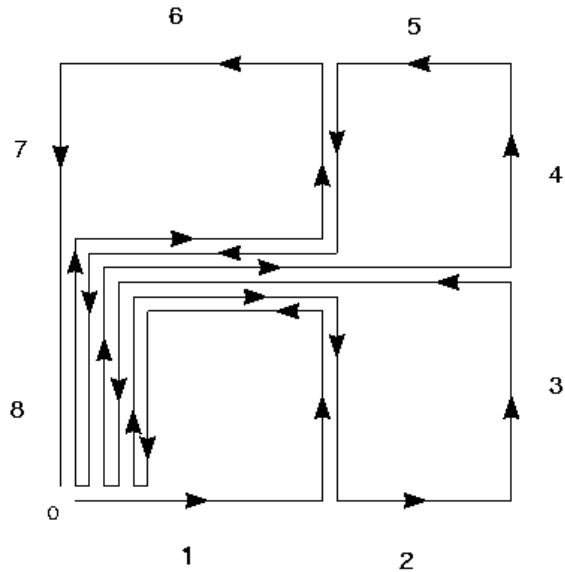}
\caption{\label{sthm1} A step in the decomposition of a simple Wilson loop into plaquettes.  [From \cite{co96}.]}
\end{figure}

\begin{figure}
\includegraphics[width=3.5in]{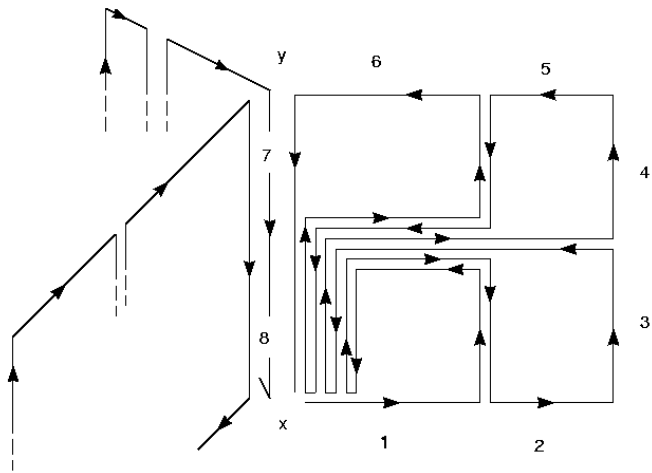}
\caption{\label{sthm2} Making plaquettes in the baryonic Wilson loop.  [From \cite{co96}.]}
\end{figure}

Possible path dependence can be understood from the non-Abelian Stokes' theorem for baryonic Wilson loops \cite{co96}.  As explained there, the appropriate surface to integrate over, for $SU(3)$, is three-bladed.  Fig. \ref{sthm1} shows an intermediate step in the standard \cite{nast} conversion of a non-Abelian contour integral into a surface integral and Fig. \ref{sthm2} shows how this loop is changed into a surface by using the construction of Fig. \ref{sthm1} on all three blades of a baryonic Wilson loop.  The central line in Fig. \ref{sthm2}, marked 78, gives a factor which can be \cite{co96} reduced to $\det U(78)=1$. Because Stokes' theorem expresses a line integral as a surface integral, it is necessary that the surface form of the integral not depend on the detailed form of the spanning surface, as long as it spans the Wilson loop contour and has no other boundaries.  This will be true in our center-vortex picture because of quantization of the magnetic flux carried by the center vortices. 

  As \cite{co96} shows, for a given center vortex of the form in Eqs. (\ref{vform}, \ref{abelpot}) the baryonic Wilson loop is, from Eq. (\ref{loop}):
\begin{equation}
\label{bwl}
W=\frac{1}{3}\exp [\frac{2\pi i}{3}\{\Lambda (1)+\Lambda (2)-2\Lambda (3)\}]+{\rm cycl.\;perm.}
\end{equation} 

\begin{figure}
\includegraphics[height=1.8in]{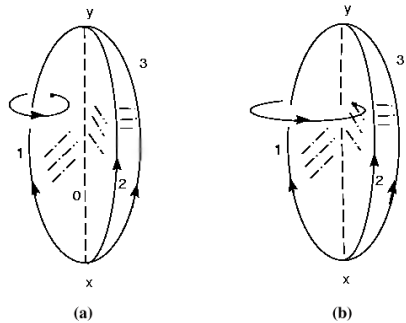}
\caption{\label{links}  (a) A center vortex (in $d=3$) linked to quark 1, and piercing the 01 surface. (b)  The same topology, with the vortex piercing surfaces 02 and 03.  [From \cite{co96}.]}
\end{figure}

We now deviate slightly from the path of \cite{co96}, where the development of Eqs. (28-33) of this reference are in error.  Evidently the $\Lambda (i)$ are not integers, although the difference of any two of them is, and it is useful to express the contents of the curly brackets in Eq. (\ref{bwl}) as differences of integers.  Consider Fig. \ref{links}(a), showing (in $d=3$, for clarity) a center vortex linked to quark line 1.  This linkage is expressed as a (signed) intersection by the vortex of a surface which we will call $\Gamma(01)$, which spans quark line 1 and the central line, labeled 0.  The link number, denoted $Lk(01)$, is the sum of intersection numbers; as shown in the figure, the link number is 1.  As mentioned above, the line integral of the Wilson loop must be independent of the position of the central line, and so precisely the same topology must be expressed in Fig. \ref{links}(b), where $Lk(01)=0,\;Lk(02)=Lk(03)=-1$, leading to a total of $Lk(02)+Lk(03)$=-2.  Although the link numbers for these configurations are different, they are equal mod(3), and this is all that counts for calculating the Wilson loop, as we will soon see. With these link numbers we can express differences of the $\Lambda (i)$ as differences of link numbers.  For example, $\Lambda (1)-\Lambda (2)=Lk(01)-Lk(02)$, since either expression is the link number of the closed loop $1\bar{2}$, in which line 2 is traced in the opposite direction.  This link number is found by summing the signed intersection number of a vortex with this loop $1\bar{2}$.  So then we express the Wilson loop phase in terms of integer differences:
\begin{equation}
\label{ldiff}
\Lambda (1)+\Lambda (2)-2\Lambda (3)=Lk(01)+Lk(02)-2Lk(03)\equiv [Lk(01)+Lk(02)+Lk(03)]\; {\rm mod} (3).
\end{equation}

We are almost through.  From Eqs. (\ref{bwl},\ref{ldiff}) the Wilson loop VEV is:
\begin{equation}
\label{bwl2}
\langle W \rangle = \langle \exp [\frac{2\pi i}{3}\{Lk(01)+Lk(02)+Lk(03)\}]\rangle.
\end{equation}
We claim that each of the phase factors is statistically independent of the other two, as simple geometrical arguments reveal, so that:
\begin{equation}
\label{bwl3}
\langle W \rangle =\langle \exp [\frac{2\pi i}{3}Lk(01)]\rangle \langle \exp [\frac{2\pi i}{3}Lk(02)]\rangle \langle \exp [\frac{2\pi i}{3}Lk(01)]\rangle .
\end{equation} 
But each of the factors in brackets is itself the expectation value of a $q\bar{q}$ Wilson loop, where, for example, the 01 factor involves the Wilson loop formed from quark line 1 and the central line 0 in Stokes' theorem (taken to be directed from $y$ to $x$).  As such, by previous arguments \cite{co03}, each bracketed factor is given by:
\begin{equation}
\label{qqloop}
\langle \exp [\frac{2\pi i}{3}Lk(01)]\rangle = \exp [-K_FA_{min}(01)]
\end{equation}
where $A_{min}(01)$ is the minimum area associated with the 01 loop.  So the baryonic Wilson loop is just:
\begin{equation}
\label{answer}
\langle W \rangle =\exp [-K_F\{A_{min}(01)+A_{min}(02)+A_{min}(03)\}].
\end{equation}
It only remains to point out that the central line 0, so far undetermined, must be a line which minimizes the sum of areas in Eq. (\ref{answer}), that is, a Steiner line.  And so we find the $Y$ area law instead of the $\Delta$ law.   

\section{Extension to   SU(N)}
\label{sun}

\begin{figure}
\includegraphics[width=3.5in]{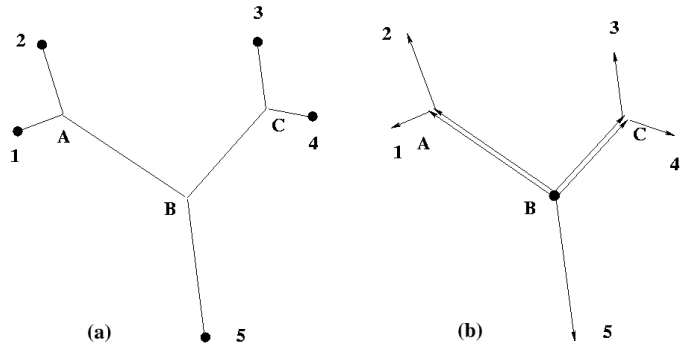}
\caption{\label{su5} (a) A view from above of a heavy-quark baryon in $SU(5)$; the quark lines are perpendicular to the figure and intersect it in the heavy dots.  Solid lines show a projection of the extremal surface spanning the quark lines.  (b) The deformation of the 5 sheets used for Stokes' theorem in $SU(5)$ to correspond to the extremal area.
The black circle at vertex B indicates the place where the 5 sheets meet in Stokes' theorem, analogous to the 78 line of Fig. \ref{su3}.}
\end{figure}

 Stokes' theorem for these higher $N$ is the obvious generalization \cite{co96} of the $SU(3)$ theorem.  We construct $N$ sheets, each bounded by a quark world line and a central line, and form plaquettes on them as we did for $SU(3)$.  This construction must be reconciled with the requirement that the baryonic area must be extremal.  It is possible to find a local minimal area consisting of these $N$ sheets meeting along a single line, by appropriate choice of the line and sheet areas; the minimum is with respect to all areas spanning the quark lines and the mid-line.  If this is the appropriate extremal area, then the Wilson-loop VEV is of the form $\exp [-K_F\sum A_{min}(0i)]$, generalizing Eq. (\ref{answer}).  However, the global minimum is constructed from surfaces joined by Steiner lines.  An example is shown in Fig. \ref{su5}(a) for $N=5$.   Fig. \ref{su5}(b) shows the deformation of the 5-bladed Stokes' theorem surface to conform to the extremal area.  Note that the surfaces whose projection is the line AB or the line BC are double, so their contribution to the phase factor in the Wilson loop for a given vortex is doubled.  That is, if a vortex of magnetic flux $2\pi J/5$, for $J=\pm 1,\pm 2$ is linked to quark line 1, as expressed in piercing the surface 1A of Fig. \ref{su5}(b), the phase factor is $\exp [2\pi iJ/5]$.    The same topology is expressed through piercings (with opposite signs) of surfaces 5B, BC, and 2A.  The piercing is counted twice for surface BC, so the alternative phase factor is $\exp [-4\pi iJ/5]$, which is equal to the first phase factor.   Unfortunately, in this case it is not possible to give an explicit formula for the Wilson-loop VEV, since in $SU(N)$  there are [$N/2$] different vortices, each with an anti-vortex, where [$n$] is the integer part of $n$ and in general, these vortices occur with different areal densities.  The baryonic Wilson-loop formula then is connected to the problem of so-called k-string tensions (string tensions for antisymmetric representations formed from the product of $k$ fundamental representations) \cite{kstring}; we will discuss it further elsewhere. 

\section{Conclusions}
\label{finis}

We have corrected an error in \cite{co96}, leading to the conclusion that the $SU(3)$ baryonic area law is the $Y$ law, not the $\Delta$ law.  The error lay in an incomplete specification of the link numbers $Lk(0i)$, and an improper specification of the statistical independence of these link numbers.

Our present result can be extended in principle to other $SU(N)$ gauge groups, at least formally, although the final result may depend on unknown areal densities of the several different types of center vortices that occur for $N>3$.  This extension uses the baryonic Stokes' theorem for larger $N$, where $N$ sheets span the quark world lines and a central line.  There is a local minimum area for such a configuration, and a simple baryonic area law which generalizes the $Y$ law in $SU(3)$.   It is also possible to deform this $N$-bladed surface to become the global minimal area spanning the $N$ quark world lines in a baryon, with the aid of Steiner lines; no explicit form for the baryonic area law is known at present.


\begin{thebibliography}{99}
\bibitem{co96}  John M. Cornwall, Phys. Rev. D {\bf 54}, 6527 (1996). 
\bibitem{co03}  John M. Cornwall, arXiv:  hep-th/0304182.
\bibitem{co77}  John M. Cornwall, Nucl. Phys. B{\bf 128}, 75 (1977).
\bibitem{suga}  T. T. Takahashi, H. Matsufuru, Y. Nemoto, and H. Suganuma, 
Phys. Rev. Letters {\bf 86}, 18 (2001); Phys. Rev. D{\bf 65}, 114509 (2002).
\bibitem{afj} C. Alexandrou, P. de Forcrand, and O. Jahn, arXiv:  hep-lat/0209062.
\bibitem{ibss}  H. Ichie, V.Bornyakov, T. Streuer, and G. Schierholz, arXiv:  hep-lat/0212036.
\bibitem{aft} C. Alexandrou, P.de Forcrand, and A. Tsapalis, Phys. Rev. D {\bf 65}, 054503 (2002); Nucl. Phys. Proc. Suppl. {\bf 106}, 403 (2002); {\bf 109}, 153 (2002).
\bibitem{nast}M. B. Halpern, Phys. Rev. D {\bf 19}, 517 (1979); N. E. Brali\'c, Phys. Rev. D {\bf 22}, 3090 (1980); I. Ya. Aref'eva, Theor. Math. Phys. {\bf 43}, 353 (1980); R. P. Feynman, Nucl. Phys. {\bf B188}, 479 (1981), P. Fishbane, S. Gasiorowicz, and P. Kaus, Phys. Rev. D {\bf 24}, 2324 (1981).
\bibitem{kstring} See, for example, L.~Del Debbio, H.~Panagopoulos, P.~Rossi and E.~Vicari,
JHEP {\bf 0201}, 009 (2002).
 

\end{thebibliography}
\end{document}